\newcommand{\boldsigma}{\mbox{\boldmath $\sigma$}}
\newcommand{\boldtau}{\mbox{\boldmath $\tau$}}
\newcommand{\bftau}{\mbox{\boldmath $\tau$}}

\newcommand{\bfP}{{\bf P}}

\def\bfk{{\bf k}}
\def\bfp{{\bf p}}  
\newcommand{\bfkap}{\mbox{\boldmath $\kappa$}} 
\def\bfr{{\bf r}}

\def\be{\begin{equation}}
 \def \ee{\end{equation}}
\def\bea{\begin{eqnarray}}
  \def\eea{\end{eqnarray}}

\newcommand{\bb}{\langle}
\newcommand{\kk}{\rangle}

\documentstyle[12pt,aps]{revtex}
\tighten
\title{
\begin{flushright}{\normalsize NT@UW-03-001}\end{flushright}
Nuclear Spin-Isospin  Correlations, \\
Parity Violation,   and the $f_\pi$ Problem}
\author{Gerald A. Miller\\
Department of Physics\\
University of Washington\\
Seattle, WA 98195-1560}

\begin{document}
\begin{titlepage}
\maketitle
\begin{abstract}
The strong interaction  effects of 
isospin- and spin-dependent 
 nucleon-nucleon correlations observed in many-body calculations
are interpreted in terms of a one-pion 
 exchange mechanism. Including such effects in computations of 
 nuclear
 parity violating effects leads to enhancements of  about 10\%. 
A larger effect arises from         the one-boson
exchange nature of 
the parity non-conserving   nucleon-nucleon interaction,  which 
 depends on 
both weak and strong meson-nucleon coupling constants.
 Using  values of the latter that are constrained
by nucleon-nucleon phase shifts 
 leads to enhancements of  parity violation by factors close to two.
Thus  
 much of previously noticed discrepancies between weak coupling constants
extracted from different experiments can be removed.
\end{abstract}
\vfill
\end{titlepage}

The problem of determining the parity-violating interaction between
nucleons has drawn recent attention with  measurements of the $^{133}$Cs
anapole moment\cite{Wood:zq}, the 
 TRIUMF measurement of proton-proton scattering\cite{Berdoz:2001nu}
and the interactions of epithermal neutrons with heavy nuclei\cite{Smith:2001db}.
The  parity non-conserving  NN potential,
 $V^{\rm PNC},$ of Desplanques, Donoghue and Holstein (DDH)
\cite{DESP80}, constructed so that the only unknown quantities
 are the weak meson-nucleon coupling constants, which govern the strength of
the weak meson-nucleon vertex function, has been the
standard tool to analyze these data because 
 the parity-violating observables have been expressed 
as linear combinations of these, presumably fundamental,
 weak coupling constants.
As explained in various reviews 
\cite{ADEL85,Haeberli:1995uz,deplanques98,Haxton:2001ay,Haxton:1997rd},
a problem occurs because 
different values of the weak coupling constants are needed to describe
different experiments.
In particular, 
the $^{133}$Cs  anapole moment  requires a pion-nucleon weak coupling constant
 $f_\pi$ that is  larger than that predicted in Ref.~\cite{DESP80},
 but the observation of circularly polarized $\gamma$ rays in the decay of 
 ${}^{18}$F
data require a very small value for $f_\pi$. Furthermore, a recent
analysis of heavy compound nuclei\cite{Tomsovic:1999yb}
finds parity violating effects that are larger (by factors $\sim1.7-3$) 
than those predicted using  DDH potential.

We next explain  why the results of nuclear structure
calculations \cite{Akmal:1997ft,SWP92} cause us to examine 
the effects of spin-isospin, nucleon-nucleon, correlations on parity violation.
Nuclear parity violating effects have been  typically analyzed in terms 
of a parity-violating single nucleon shell model potential,  constructed
from the DDH potential using a Hartree-Fock approximation 
and RPA correlations 
({\em e.g.} \cite{ADEL85,Tomsovic:1999yb,Flambaum:1997um}).
 However, two-particle-two-hole correlations are known to be as  vital elements
of nuclear structure. The spin-independent effects of the short
range correlations are often incorporated \cite{ADEL85} using the
Miller-Spencer correlation function\cite{Miller:1975hu}, 
but this does not take into account all
of the correlation effects\cite{gari,deplanques98}.

In particular, recent variational 
studies of nuclear matter \cite{Akmal:1997ft}
and $^{16}$O \cite{SWP92} have demonstrated that
spin-isospin correlations are very important. To be specific, consider
 two-nucleon pair correlation functions defined by the 
expectation values 
\bea 
\rho_\Gamma(r)={1\over4\pi r^2 A}\bb\Psi\vert \sum_{i<j}\delta(r-r_{ij})
\Gamma_{i,j}\vert\Psi\kk,\label{pair}\eea
with $r_{ij}=\vert \bfr_i-\bfr_j\vert,\;$ and $\Gamma_{i,j}$ are 
various two-nucleon operators. For the central term  $\Gamma_c=1$. 
The work of Refs.~\cite{Akmal:1997ft,SWP92} 
found that with 
$\Gamma_{i,j}=\bftau_i\cdot\bftau_j\;\boldsigma_i\cdot\boldsigma_j,\;
(\Gamma=\sigma\tau)$ or 
$\Gamma_{i,j}=\bftau_i\cdot\bftau_jS_{ij}\; (\Gamma={\rm ten}\tau)$, the
$\rho_\Gamma(r)$ can be relatively large,
even  larger than  the well-known effects of  short-distance
repulsion.  
Other operators $\Gamma$ cause  much smaller effects  and are ignored here.

 What is the impact of  spin-isospin correlations
 on calculations of parity-violating observables? Consider 
the construction of the single-particle PNC potential
 $\widehat{U}^{\rm PNC}$ which
governs the interaction between  a valence nucleon  $(i)$ with   
 a spin 0$^+,\;N=Z$ core.
In using the 
Hartree-Fock approximation, one folds the operator $V^{\rm PNC}$
with the density matrix of the core. 
The pion exchange term contains the operator
$i({\boldtau_i\times\boldtau_j})_z$, where $j$ represents the core nucleons.
The expectation value of this operator vanishes.
 But if one includes spin-isospin correlations, the relevant matrix elements
include a factor $ \bftau_i\cdot\bftau_j$, and the product 
$i({\boldtau_i\times\boldtau_j})_z\bftau_i\cdot\bftau_j=-2
(\bftau_i-\bftau_j)_z-i({\boldtau_i\times\boldtau_j})_z$ contains a 
non-vanishing term, $-2\bftau_i$.
Thus a new non-vanishing contribution will appear.  

The starting point for our evaluation of the PNC single-particle 
potential is the
effective parity-violating nucleon-nucleon potential\cite{DESP80} between
two nucleons (i,j):
\begin{eqnarray}
&&MV^{\rm PNC}(i,j)=i{f_\pi g_{\pi NN}\over \sqrt{2}}
\left({\boldtau_i\times\boldtau_j\over2}\right)_z(\boldsigma_i+\boldsigma_j)\cdot
{\bf u}_\pi(\bfr)
\nonumber\\
&&-g_\rho \left(h_\rho^0\boldtau_i\cdot\boldtau_j+h_\rho^1\left({\tau_i+\tau_j\over 2}\right)_z
+h_\rho^2{(3\tau_i^z\tau_j^z-\boldtau_i\cdot\boldtau_j)\over 2\sqrt{6}}\right)
\nonumber\\
&&\times\left((\boldsigma_i-\boldsigma_j)\cdot
{\bf v}_\rho(\bfr)
+i(1+\chi_V)
\left(\boldsigma_i\times\boldsigma_j\right)\cdot
{\bf u}_\rho(\bfr)\right)
-g_\omega\left(h_\omega^0+h_\omega^1+\left({\tau_i+\tau_j\over 2}\right)_z\right)\nonumber\\
&&\times\left((\boldsigma_i-\boldsigma_j)\cdot
{\bf v}_\omega(\bfr)
+i(1+\chi_S)\left(\boldsigma_i\times\boldsigma_j\right)\cdot
{\bf u}_\omega(\bfr)\right)
-(g_\omega h_\omega^1-g_\rho h_\rho^1)
\left({\tau_i-\tau_j\over 2}\right)_z\nonumber\\
&&\times(\boldsigma_i+\boldsigma_j)\cdot
{\bf v}_\omega(\bfr)
-g_\rho h_\rho^{1'}i\left({\tau_i\times\tau_j\over 2}\right)_z
(\boldsigma_i+\boldsigma_j)\cdot
{\bf u}_\omega(\bfr)
\label{DDH}\end{eqnarray}
where $M$ is the nucleon mass, 
${\bf v}_m(\bfr)\equiv \left\{\bfp,f_m(r)\right\},\;
{\bf u}_m(\bfr)\equiv \left[\bfp,f_m(r)\right],\;$
$f_m(r)=\exp (-m_mr)/4\pi r$ (with $m_m=m_{\pi,\rho,\omega})$. 
 The 
strong interaction parameters used by DDH  are 
$g_{\pi NN}=13.45,\;g_{\rho NN}=2.79,g_{\omega NN}=8.37,\; 
\chi_V=3.7,$ and $\chi_S=-0.12$.
The formula (\ref{DDH})  is  still            used,
\cite{Tomsovic:1999yb,Haxton:2001zq}-
\cite{Liu:2002bk} with these original strong interaction parameters.

The effect we wish to incorporate is that Eq.~(\ref{DDH}) is not the
complete PNC interaction  between two-nucleons. This is because
the  PNC potential acting once,  occurs in the midst of all orders of the
strong potential,  $V$.
 The resulting PNC reaction matrix $G^{\rm PNC}(E)$
 is a generalization of   the Bruckner reaction matrix and is given by:
\bea
&&G^{\rm PNC}(E)= \Omega^\dagger(E) V^{\rm PNC}\Omega(E),\label{2pt}\\
&&\Omega(E)=1 +{Q\over E-H_0}V\Omega(E)
=1 +{Q\over E-H_0}G(E),\label{om}\eea
where $E$ is an energy to be discussed below, and $Q$ is an 
operator which projects onto unoccupied states.

Our focus is on a first estimate of the effects of  spin-isospin correlations.
Thus we consider a nucleon of momentum $\bfP$ outside a core 
which is approximated as infinite nuclear matter. In this case, 
 the PV single particle
potential $\widehat{U}^{\rm PNC}$ is given by:
\bea
\bb s,t\vert  \widehat{U}^{\rm PNC}(\bfP)\vert s',t'\kk=
\sum_{\bfk,(k\le k_F)m,m_t}\bb\bfP,s,t;\bfk,m,t\vert 
G^{\rm PNC}(E)\vert\bfP,s',t';\bfk,m,m_t\kk_A. \label{pvu}
\eea

The goal of this work is to assess the influence of 
tensor and spin dependent correlation
 effects on calculations of PNC observables. Since 
these are usually ignored, performing a schematic 
 calculation seems appropriate. 
It might seem easiest to parameterize the
different contributions to $\Omega(E)$ as function of $r$,  the
procedure of  Miller \&
Spencer. Such a strategy will not work here, with the emphasis  on finding
hitherto neglected contributions to the 
 direct term. To see this consider the terms of the DDH potential of the
form 
$[\bfp,f_m]\sim  \boldsigma\cdot \bfr$.
 The Miller-Spencer procedure would be
 to treat
$\bb\bfr\vert \Omega(E)\vert \bfkap,m,m_t\kk$ as a function of $r$,
so that  the direct matrix element of
Eq.~(\ref{pvu}) would involve a vanishing 
volume integral of $\boldsigma\cdot \bfr$ 
times a function of $r$. Indeed,
 the
 PV effective potential arises from  the  dependence of $\Omega(E)$
on the relative
momentum 
\bea \bfkap =(\bfP- \bfk)/2.\eea

The importance of the spin-isospin correlations arises from the exchange
of pions \cite{Akmal:1997ft,SWP92}, so we separate  the  potential $V$
into a one pion exchange term, $V_{\rm OPE}$ and a remainder 
approximated as being a central potential, $V_c$.  We include the 
effects of the
central potential (which include the short-distance  repulsion) to all orders
and keeps terms of first-order in   $V_{\rm OPE}$. The application of the 
 two-potential theorem to Eq.~(\ref{om}) gives
\bea 
\Omega(E) \approx\Omega_c(E)+\Omega_c^\dagger(E){Q\over E-H_0}V_{\rm OPE}\;
\Omega_c(E),\label{s2pt} \eea
with $\Omega_c(E)\equiv 1+{Q\over E-H_0}V_c\;\Omega_c(E),$ and $Q$ is the
usual operator which projects onto two-particle-two-hole states. 
It is natural to model
$\Omega_c(E)$ as 
\bea
(\bfr\vert\Omega_c(E)\vert\bfkap)= (1+f_c(r))e^{i\bfkap\cdot\bfr},\eea  so that
Eq.~(\ref{s2pt}) can be expressed as
\bea (\bfr\vert\Omega(E)\vert\bfkap)=(1+f_c(r))e^{i\bfkap\cdot\bfr}
+{\hat\psi}_{\bfkap}(\bfr)\label{ope},\eea with
\bea\widehat{\psi}_{\bfkap}(\bfr)=
(1+f_c(r))\int{d^3p\over (2\pi)^3}{M\over p^2+\omega^2}
e^{i\bfp\cdot\bfr}(\bfp\vert\;V_{\rm OPE}(1+f_c)\vert\bfkap).\eea
The notation $\vert )$ denotes   a spatial overlap,  so that 
 $\widehat{\psi}_{\bfkap}
(\bfr)$ are operators in spin-isospin space. 
The parameter $\omega\;(\omega^2/M=-E)$, will be 
chosen to reproduce the results of Ref.~\cite{Akmal:1997ft}. Using a 
 negative value
of $E$ causes the two-particle-two-hole fluctuations
to have a short range of order $\sim1/\omega$. This 
allows us to neglect the effects of the  operator $Q$ because 
 terms involving $1-Q$
can be regarded as correction terms,  of higher order in the density
than the terms we examine here. 

The main result of this analysis is the simple model (\ref{ope}). 
It is necessary to compute 
the matrix elements of the pair-correlation operators of Eq.~(\ref{pair}) to
see if the results of Ref.~\cite{Akmal:1997ft} can be reproduced with such 
a simple formula.  Evaluating Eq.~(\ref{pair})
keeping only two-nucleon correlations leads to  the result:
\bea
&&\rho_\Gamma(r)=\sum_{S,M_S,T,M_T}\int\;{d^3\kappa\over (2\pi)^3}
\theta (k_f-\kappa)f(\kappa)\bb\bfkap,S,M_S,T,M_T\vert \widehat{\rho}_\Gamma(r)
\vert\bfkap,S,M_S,T,M_T\kk_A,\label{rhoval}\\
&&\widehat{\rho}_\Gamma(r)\equiv{2\over 4\pi r^2} \Omega^\dagger(E)\;
\sum_{i<j}\delta(r-r_{ij})\Gamma_{ij}\;\Omega(E),\eea
where
$f(\kappa)=1-{3\over 2}({\kappa\over k_F})+{1\over 2}({\kappa\over k_F})^3$
and
$\vert\bfkap,S,M_S,T,M_T\kk_A\equiv \vert\bfkap,S,M_S,T,M_T\kk-(-1)^{S+T}
\vert-\bfkap,S,M_S,T,M_T\kk$.
We take $k_F=1.36\;{\rm fm}^{-1}$ and choose the previously determined
\cite{Miller:1975hu}
function 
 $f_c(r)=-e^{-\alpha r^2}(1-\beta r^2)$ with
$\alpha=1.1\;{\rm fm}^{-2},\;\beta=0.68\;{\rm fm}^{-2}$.
The value of  $\omega$ that leads to  reproducing  the results of
Ref.~\cite{Akmal:1997ft} is
$\omega=2fm^{-1}$,  which is 
of the expected order $\sim        k_F$.
As shown in 
Fig.~1, and the present results  reproduce the qualitative features
of the pair-correlation functions  of Ref.~\cite{Akmal:1997ft}.
Furthermore,  the integrals of our $\rho_\Gamma$ are in quantitative 
agreement with those of Ref.~\cite{Akmal:1997ft}, so the present model
should provide an adequate first assessment of the impact of such
  correlations on calculations of PNC observables.
\begin{figure}
\protect
  \unitlength1.8cm
\begin{picture}(6,6)(-0,-6.2)
\includegraphics{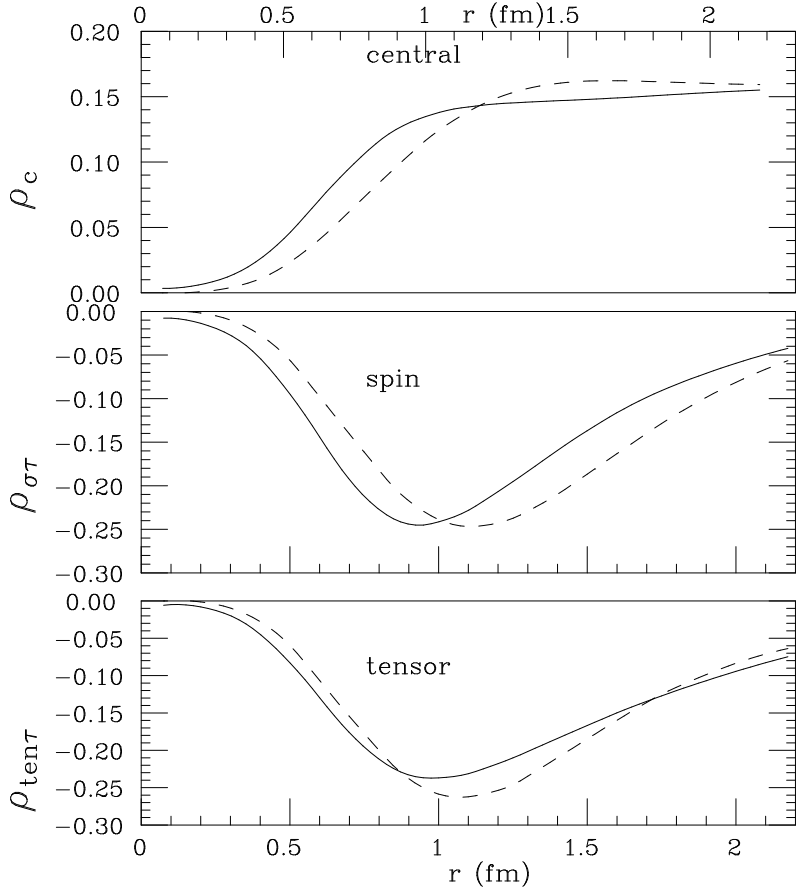}
\end{picture}\vspace{1.7cm}
\caption{\label{figcor}\protect Pair-correlation functions. Solid curves are
from Akmal \& Pandharipande [11], dashed curves are from this work.
  }
\end{figure}

Given our model of Eq.~(\ref{ope}), we can now 
 evaluate the operator   $\widehat{U}^{\rm PNC}$ of
Eq.~(\ref{pvu}). Using  Eq.~(\ref{ope}) in Eq.~(\ref{2pt}),
and keeping  terms of first order in $V_{\rm OPE}$ leads to
\bea
G^{\rm PNC}(E     )&\approx&            \widetilde{V}^{\rm PNC}+
\Omega_c^\dagger(E)V_{\rm OPE}{Q\over E-H_0}\widetilde{V}^{\rm PNC}
+\tilde{V}^{\rm PNC}{Q\over E-H_0}V_{\rm OPE}\Omega_c(E),\label{gpnc}\\
&=&\widetilde{V}^{\rm PNC}+\Delta G^{\rm PNC}(E     ),\label{2term}
\eea
where 
$\widetilde{V}^{\rm PNC}\equiv(1+f_c){V}^{\rm PNC}(1+f_c).$
The use of Eq.~(\ref{gpnc}) in Eq.~(\ref{pvu}) specifies our model\cite{apr}.
The numerical evaluations are straightforward, so we 
present the results.

The operator  $\widehat{U}^{\rm PNC}(\bfP)$ of Eq.~ (\ref{pvu}) must 
be  a pseudo-scalar spin-isospin operator. For our nuclear matter problem 
it is of the form
\bea \widehat{U}^{\rm PNC}(\bfP) =\boldsigma\cdot 
\bfP{\rho\over M^3} (g_0+g_1\tau_z).\label{note}\eea 
For $^{133}$Cs and $^{205}$Th, $(N-Z)/(N+Z)\approx 0.17$, and we find that
 keeping the term $g_1$ amounts to keeping a correction to a term that
is not large. So we take 
$N=Z$ causing  $g_1$ to vanish. To assess the importance of the spin-isospin
correlation, we compare the influence of the two terms of Eq.~(\ref{2term})
in Eq.~(\ref{pvu}). Denoting the results of using the first ($V^{\rm PNC}$)
term as  $g_0^{(0)}$
and those of using the second term as $\Delta g_0  $, 
with $g_0=g_0^{(0)}+\Delta g_0$, and evaluating the matrix element numerically
leads to the results:
\bea
g_0^{0}&=&24\;f_\pi-6.9 \;h_\rho^0-3.6\;h_\rho^1-4.1 \;h_\omega^0
-4.1 \;h_\omega^1 
\\
\Delta g_0&=&2.43\;f_\pi -1.2 \;h_\rho^0-.66\;h_\rho^1 -.351 \;h_\omega^0
-.15   \;h_\omega^1 
\eea
We see that the coefficient of each term is enhanced by about $10\%$
if $\Delta g_0$ is included.  This is in contrast
with many other nuclear structure effects which reduce the effects of 
PNC \cite{Haxton:2001ay,Haxton:1997rd}. The present results 
constitute an argument that the effects of spin-isospin correlations need
to be included in quantitative calculations of PNC effects, but 
are not large enough to not have
a major impact\cite{smspace}.  Indeed, uncertainties due to nuclear structure
effects might not have a large impact on  PNC observables.

But there is a more obvious source of change to present 
 evaluations. Examination of 
the DDH potential, Eq.~(\ref{DDH}) immediately reveals 
the dependence on 
 the product of strong and weak coupling constants. In a 
one-boson-exchange model one uses one strong and one 
weak meson-nucleon vertex function to  construct the 
potential. But the strong coupling constants can be separately determined
by computing the  potential $V$ and choosing the parameters to reproduce 
experimentally measured scattering observables. This is the procedure of
{\it e. g.} the Bonn potential\cite{Machleidt:hj}. As shown in Table I, 
the strong coupling constants
required to fit phase shifts are much larger than those used originally by DDH
(and used presently in Refs.\cite{Haxton:2001zq,Schiavilla:2002uc,Liu:2002bk}).
\begin{table}
 \begin{center}
\begin{tabular}{|c|c|c|}
\hline
\quad   & DDH & Bonn (OBEPR) \\ \hline
$g_{\pi NN}$ & $13.45         $ &13.68 \\
$g_\rho$& $2.79             $&3.46\\
$\chi_V$& $3.7            $& 6.1\\
$g_\omega$& $8.37              $&20\\
$\chi_S$&$ -0.12   $& 0.0      \\
\hline\end{tabular}
\caption {Comparison of strong coupling constants.} 
\end{center}
\end{table}
The coordinate-space 
potential (Table 14) of Ref.~\cite{Machleidt:hj} is used for this comparison 
because it is a non-relativistic, local potential that 
 is technically compatible with
Eq.~(\ref{DDH}).     The large values of $\chi_V$ and $g_\omega$ are essential
requirements to reproduce data, if a one-boson-exchange model is used
\cite{Machleidt:tm}.

Now consider the impact of using the Bonn parameters of Table I, in the
DDH potential.  If one considers a nucleon outside an N=Z core,
the parameter combinations $h_\rho^{0,1}g_\rho(1+\chi_V)$ and $
h_\omega^{0,1}g_\omega$
determine the vector meson contributions to 
$U^{\rm PNC}$\cite{ADEL85}. Thus using the Bonn parameters 
is equivalent to increasing the coefficient of the terms proportional to
$h_{\omega,\rho}^{0,1}$ by a factor of 1.9.  The effects of using values
such as those of Table I have been noticed 
previously\cite{Nessi-Tedaldi:ak,Driscoll:1988hg}.

\begin{figure}
\protect
  \unitlength1.8cm
\begin{picture}(6,6)(-0,-6.2)
\includegraphics{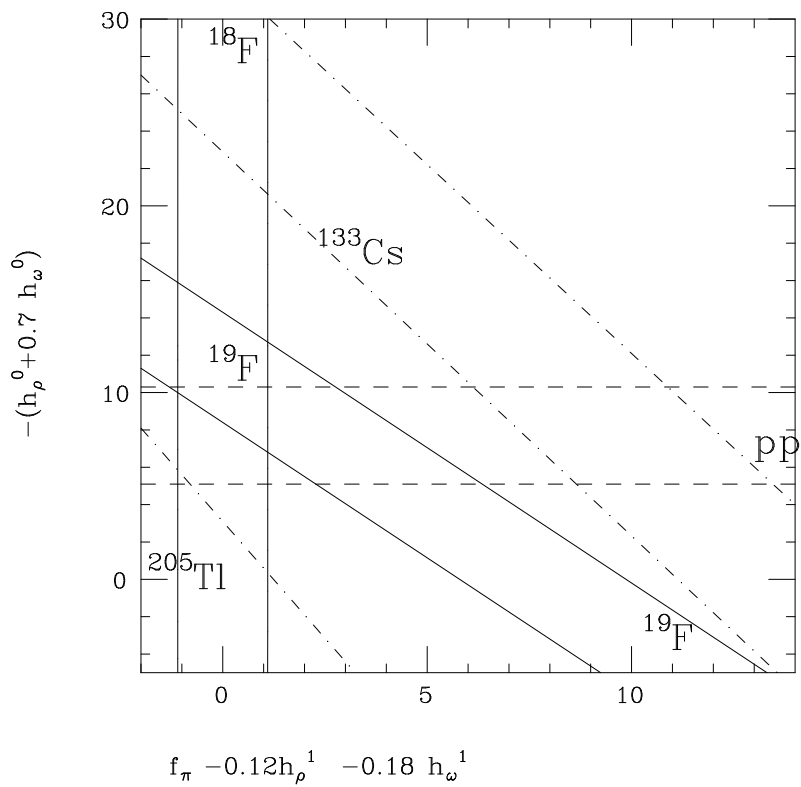}
\end{picture}\vspace{1.7cm}
\caption{\label{figdat} Constraints on the PNC coupling constants $(\times 10^7)$
that follow from  using 
the Bonn coupling constants of Table I.} 
\end{figure}
To  make a first assessment of the impact of this finding, we revise the 
results of Table VII and Fig.~9 of Ref.~\cite{Haxton:2001zq} 
by multiplying the
coefficient of the terms proportional to $h_{\omega,\rho}^{0,1}$
 by a factor of 1.9\cite{quibble}.  This leads to Fig.~2. The allowed regions are
between the dashed lines ($pp$), each set of solid lines  ($^{18,19}F$),
and each set of dot-dashed lines $^{133}$Cs, $^{205}$Tl.
The results of the $pp$, and $^{18,19}$F experiments can be explained with
one set of weak-coupling parameters. 
The  $^{205}$Tl result with its large
error band is almost consistent with that  set of parameters, but 
the $^{133}$Cs, and $^{205}$Tl experimental 
results are not compatible, as pointed out earlier\cite{Haxton:2001zq}.
The results from compound nuclei are typically not shown in plots such as
 Fig.~2, but using the Bonn parameters of Table I, 
would reduce the discrepancies between theory and experiment.

The constraints arising from  $(p,\alpha)$ scattering
experiments are not shown in Fig.~2, because 
 currently interpreted provides constraints very similar to
that of the $^{19}$F experiment, and 
 the calculations are not not complete. For example, 
 $\widehat{U}^{\rm PNC}$, which should be complex,  is treated as  real.
  But it is worth remarking that
an $(n,\alpha)$ experiment would provide very different constraints.
The parity violation predicted using the original DDH potential is 
very small due to a cancellation 
between the pionic and vector meson exchange terms\cite{Flambaum:xu}. The
 use of the Bonn coupling constants  (Table~I) leads
to a value of $g_n=1.9 $ (see Eq.~(18) of \cite{Flambaum:1997um})
instead of $g_n=0.2$. This would 
 be  increased further 
by using a smaller value of $f_\pi$ (indicated in
Fig.~2) than the DDH ``best guess value'' of  Ref.~\cite{Flambaum:1997um}.
If $f_\pi=1$, then $g_n=3.2$.

The ranges of weak-coupling constants covered by Fig.~2 are roughly consistent
with the DDH ``reasonable ranges'' and the same coupling constants account
for much of the data, except for that of $^{133}$Cs. 
This constitutes some success in explaining PNC phenomena, but the
main improvement obtained here by using larger strong coupling constants could
itself be true only within one-boson exchange models. All such models have
large values of $\chi_V$ and $g_{\omega NN}$, but two-boson exchange models
have smaller values of $g_{\omega NN}$\cite{Machleidt:hj}.
Furthermore, 
the one-boson-exchange approach is not consistent with the present state of
the art treatments of nucleon-nucleon scattering, 
see Refs.~\cite{Entem:2001cg,Epelbaum:1999dj},
and  \cite{vanKolck:1999mw}. An appropriate treatment of 
 PNC effects is one which  involves  incorporating PNC effects within
 an updated treatment of the nucleon-nucleon interaction. For example, 
Holstein
has recently advocated 
an effective field theory treatment\cite{barry}. 

Thus our summary is that the  
present results demonstrate the need for an improved, updated, 
 consistent incorporation of strong
and weak interaction effects, and also  indicate that the 
nuclear structure uncertainties might not be very severe. Thus the 
improvement of calculations of nuclear PNC effects seems to be an interesting
and feasible task, even though  much remains to be done.

\section*{Acknowledgments}
This work is partially supported by the USDOE. This work has  benefited from
conversations with W.~C.~Haxton, B.~R.~Holstein, C.~P.~Liu, O.~P.~Sushkov 
and U. van Kolck.


\begin{thebibliography}{99}
\bibitem{apr} The higher order terms lead to negligible PNC effects
and are also negligible in Eq.~(\ref{rhoval}).
\bibitem{quibble} Although $V^{\rm PNC}$ is of first-order in
$g_{\omega NN}$ and $g_{\rho NN}(1+\chi_V)$,
 the  observable quantities 
depend on these coupling constants in a non-linear
manner. This effect is ignored here.
\end{thebibliography}
\end{document}